# A Novel Channel Reservation Mechanism for Next Generation Cellular Networks

D. Sreenivasulu[1] and P. Venkata Krishna[2]

[1]Research Scholar, Department of Computer Science Rayalaseema University, Kurnool, AP, India
Email: dsnvas@gmail.com

[2]Senior Member IEEE, Department of Computer Science, Sri Padmavati Mahila University Tirupati, India
Email: dr.krishna@ieee.org

*Abstract*

*The increase in the number of mobile users increase in the requirement of spectrum. When effective and efficient channel allocation procedures are introduced, the requirement can be reduced. As the users move from one location to the other and wish to continue their communication without any interruption, handoff process need to be handled transparently such that the user is not aware of the change in the control of base stations. Hence, a deep learning-based channel reservation mechanism is proposed in this paper along with channel allocation procedure.*

*.*

## 1  Introduction

The expanding interest for the communication process in wireless networks where the spectrum resource is very less in quantity lead to the advancement of the procedure for channel allocation which behave dynamically for wireless networks which are rapidly evolving in the today's world. A virtuous outline for the dynamic channel allocation procedures is given in [4]. The communication procedures need to be refined as to support the present and upcoming societal growth towards technology. The day-by-day growth in mobile users, there will be a huge demand for the various forms of required data, which may be used professionally, personally, or for entertainment [1, 2]. As the demand of the required information increases, traffic in the network increases. Future generation networks need to support various diversified applications with high reliable, scalable, low latency, variable rates at which the transmission take place, etc. Ultra-high frequency bands are used in cellular network, which is profoundly made used, and hence make the operators to get more frequency band to service more number of users. In this regard, an effective and efficient usage of frequency band is highly required. The network becomes more densified in the coming days, which will be supported by the next generation networks [5]. The next generation networks will be capable of providing the spectrum which enable to connect more number of devices and applications together at any place and any time [6].

## 2 Deep Learning

The deep learning models [28-30]consists of feed forward neural network model and recursive neural network model. Multi-layer perceptron model and convolutional neural network model belongs to feed forward network model. Recursive Neural Networks is a type of neural network, which can be seen as a generalization of the recurrent neural network. Recursive Neural Networks have successfully applied to processing data structures as inputs to neural nets. In recursive neural networks, features are extracted and these features are represented in a high dimensional vector space and then these features are submitted to the neural network.

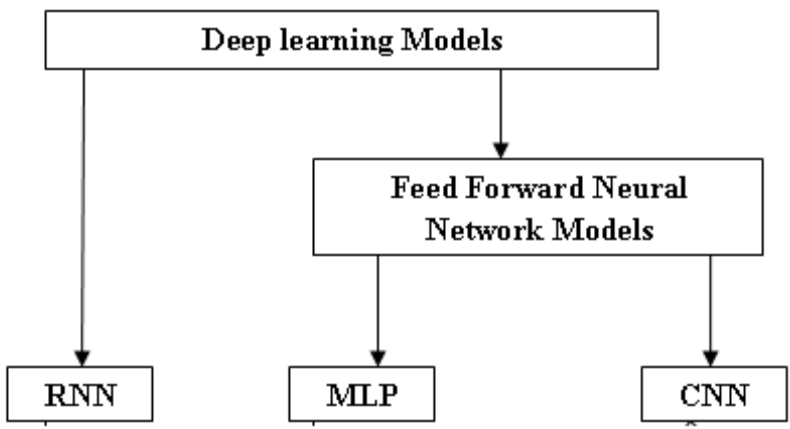

**Figure 1: Deep Learning Models**

Multi-Layer Perceptron (MLP) and Convolutional Nueral Networks (CNN) are considered as feed forward neural network models as shown in Fig. 2. MLP is a neural network model that consists of feed forward neural network with one or more hidden layers between input and output layer. Here the perceptron will act as either activation function or binary classifier. The single layer perceptron is called wide learning component, which can be viewed as a generalized linear model. The most widely used generalized linear model is a logistic regression model [28-30].

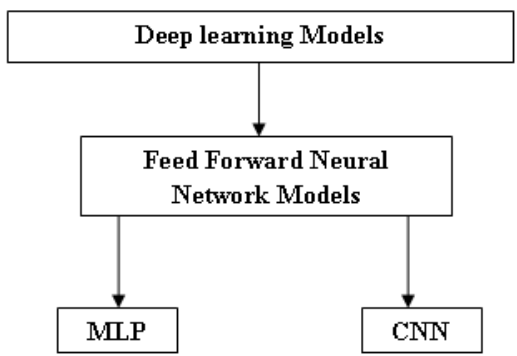

**Figure 2. Deep learning Models with Feed forward neural network.**

Many fields like manufacturing, healthcare, etc. have an impact of Deep learning. Various kinds of complex problems such as recognizing objects, speech, etc. are being solved by the industries using this deep learning concept. Deep learning is an amicable appellation for an artificial neural network. The keyword "Deep" in "Deep Learning" indicates the level of learning in the network. The complexity of the system increases the depth of the network increases. Deep learning can be considered as the learning technique using more number of layers of notion to unravel complex problems. In olden days, only single layer is being used in the learning based on neural networks due to high cost for calculations. So, it can be inferred that the deep learning systems are expensive and complex when compared to single layered neural networks. Generally, humans can recognize the tasks or picture better when compared to the machines but with the embedding of deep learning concept in computer vision applications, the scene has been reversed and humans are defeated in recognizing the things. The deep learning concept is used by Facebook to improve its performance in detecting the faces in snapshots, pictures, etc. The human is able to identify the same person from various number of photos at 97.53% whereas the machines are equally competing with 97.25% irrespective of other constraints like colour, light, direction of face, etc. hence, this enhancement of the system is not just on the edge but it can be termed as game changer. Speech recognition became more popular and is being very widely used in the mobile phones to convert the voice to the text. There are many languages and it is tedious task to convert the speech of all languages into voice. China introduced the conversion of speech into text for two languages – English and Mandarin in one of the search engine called Baidu. This search engine is quite quicker and precise in performing the tasks. Deep learning can be used in reducing the energy consumption also. Google minimized the energy consumption to 40% at their data centers by implementing deep learning concept and hence able to save hundreds of millions of dollars.

The rest of the paper is organized as follows: related work is presented in section 3, deep learning based channel reservation mechanism along with channel allocation procedure is described in section 4, performance analysis is performed in section 4, section 5 projects the results and discussion is carried out and finally section 6 concludes the paper.

## 3  Related Work

In [3], the authors discussed the overall overview of 5G structures along with probable necessities and the technology which help in accomplishing their promising objectives. The authors also presented communication systems in the view of complicated systems.

A new scheme of distributed channel allocation in cellular networks is proposed in [10]. According to this scheme, handoff calls are considered to be more preferable and hence some channels are reserved for them at the initial stage. These reserved channels for handoff calls are altered from time to time based on the performance of the system in terms of blocking probability and dropping probability. Various types of traffic like real time and non-real time traffic are considered in the proposed system. All types of calls are given more time for trying for the communication channel by maintaining queues. Three-dimensional Markov model is used to represent the system and analysis is made to evaluate dropping and blocking probabilities.

Learning automata is used in channel allocation procedure for VANET in [11]. This procedure is an enhancement of the procedure in [10]. The number of channels to be reserved for handoff calls is selected using learning automata so that the performance is optimum. Various parameters like dropping probability, blocking probability and handoff latency are used to estimate the performance of the proposed system.

Learning automata is used for reserving the channels for the handoff calls in the distributed channel allocation procedure proposed in [12]. The performance of various types of system models as single traffic no queues, single traffic with queues, multi-traffic no queues and multi-traffic with queues are evaluated. In single traffic no queues system model, no variation between real time and non-real time traffic and also queue is not maintained either for originating calls or handoff calls. In single traffic with queues, queues are maintained for both the originating and handoff calls. In multi-traffic no queues, queues are not used but real time and non-real time traffic are distinguished. In multi-traffic with queues, traffic is categorized as real and non-real time and also queues are maintained for all calls. The system performance is assessed using LA and without using LA.

A cross layer-based handover scheme is introduced by the authors in [13]. The cross layer is developed between physical layer and MAC layer in order to enable the sharing of information between these two layers. Handoff latency is considered to be the main parameter for evaluation and hence, to reduce the handoff latency, the position characteristic which can be obtained from physical layer is communicated to MAC layer on before hand so that the skim through time can be decreased. Oncoming Side Vehicles and relay nodes are used to minimize the handoff latency and enhance packet delivery ratio. The vehicles which are out of communication range of oncoming side vehicles, receive the information using relay vehicles. If the links are strongly connected, then handoff latency could be maintained low. Relay vehicles use WiMAX Multihop Relay technology to provide the service of Internet inside the vehicle. This system performs good when there are more number of oncoming side vehicles and relay nodes otherwise the system performance degrades. When more oncoming side vehicles and relay nodes are used, the system becomes expensive. Hence, the performance depends on the arrival rate of oncoming side vehicles and relay nodes and so, there is a trade-off here, which the author missed to consider in this paper.

In [14], the authors proposed an algorithm to optimize the channel allocation and power allocation in Non-Orthogonal Multiple Access (NOMA). NOMA is used to make more than one communication to use same frequency during the same time with the help of Successive Interference Cancellation. The problems are categorized into tractable and intractable. Optimal solution is provided for tractable problem and near-optimal solution is provided for intractable problems. The authors claim that the proposed system enhances the performance of the system in terms of fairness and throughput for both channel allocation procedure and multiple access orthogonally.

The authors in [15] analytically proved that the resource allocation in NOMA is NP-hard. Lagrangian duality and dynamic programming is used to provide almost optimal solution for the proposed method. The performance of the proposed algorithm is compared with orthogonal frequency division multiple access (OFDMA) and also with legacy systems for NOMA.

In [16], the authors proposed a procedure for allocating resources for enabling the communication process among any 2 vehicles which behave as nodes in the network. The communication link between the two vehicles are free enough to choose the channel and the amount of power required for communication. Hence, the system is said to be decentralized and also do not increase the overhead of communication. Deep reinforcement learning is used to reduce the interference during the communication among vehicles.

In [17], the authors devised a procedure based on deep learning concepts to demonstrate the issue of allocating resources for LTE-U small base stations. According to the proposed method, more number of small base stations are capable of choosing the channel dynamically, integrating the carrier or split the range with assurance of impartiality among prevailing networks.

## 4. Deep Learning based Channel Reservation Mechanism

The various types of traffic considered are real-time and non-real time. Real-time traffic is delay sensitive and non-real time traffic is not delay sensitive. Hence, more priority is given to real time traffic. The types of calls considered are originating calls and handoff calls. Originating call is the one which is initiated and is under the control of the present Base Station (BS). The handoff call is the on-going call, means initiated in the communication range of one BS and moving into the communication range of another BS. So, the combination of the traffic and calls is considered as real-time originating calls, non-real time originating calls, real time handoff calls and non-real-time handoff calls. When a channel is not allocated to requested origi-nating call, then it is said to be blocked and if a channel is not available to allocate to requested handoff call, it is said to be dropped. The complete process is divided into channel reservation procedure and channel allocation procedure. Initially, the number of channels that are under the control of BS do not reserve any channel for any type of calls. The system is made to learn deeply and then the number of channels that are required for various types of calls is decided. Based on this decision, the number of channels are reserved for each type of call. Periodically, this reservation is updated based on the performance of the network in terms of blocking probability, dropping probability, handoff latency. The probability with which the originating call lose the chance of acquiring a channel is called as blocking probability and the probability with which the handoff calls lose the chance of acquiring a channel is called as dropping probability. The time within which the control of a particular from one base station to other base station is referred as handoff latency. Assume that the channels reserved for various types of calls NOC, ROC, NHC, RHC. The non-real time originating calls can access only NOC channels, real-time originating calls can access only ROC channels, non-real time handoff calls can access only (NHC – ROC) channels and real time handoff calls can access only (RHC – ROC) channels. The system model depicting all these channel reservations for various types of call is shown in Fig. 3.

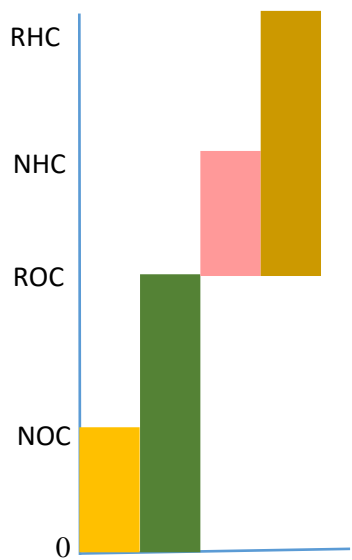

**Figure 3: Proposed System Model**

## 5  Result Analysis

The proposed algorithm is simulated and its performance is evaluated in terms of blocking probability, dropping probability and handoff latency. The performance in terms of blocking probability is shown in Fig. 4, dropping probability is shown in Fig. 5, handoff latency is shown in Fig. 6. For simulation purpose, the number of cells in the network are considered to 15, number of channels are considered to be 60, arrival rate of real time originating calls is considered to be 12 calls/s, arrival rate of non-real time originating calls is considered to be 20 calls/s, arrival rate of real time handoff calls is considered to be 5 calls/s and arrival rate of non-real time handoff calls is considered to be 10 calls/s and the service rate is considered to be dynamic based on the status of the network. The mobility of the nodes is considered to be 20 m/s.

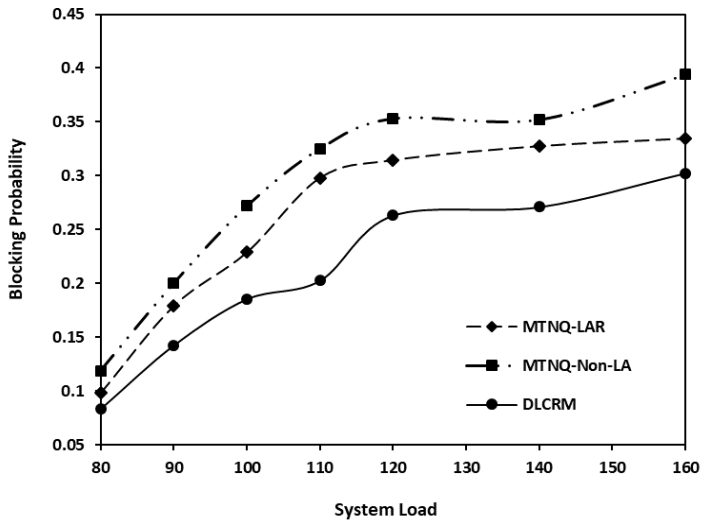

**Fig. 4 Blocking Probability vs System Load**

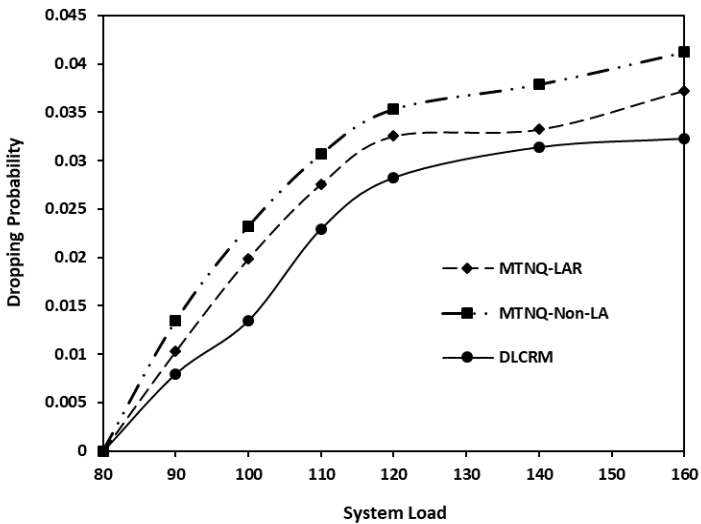

**Fig. 5 Dropping Probability vs System Load**

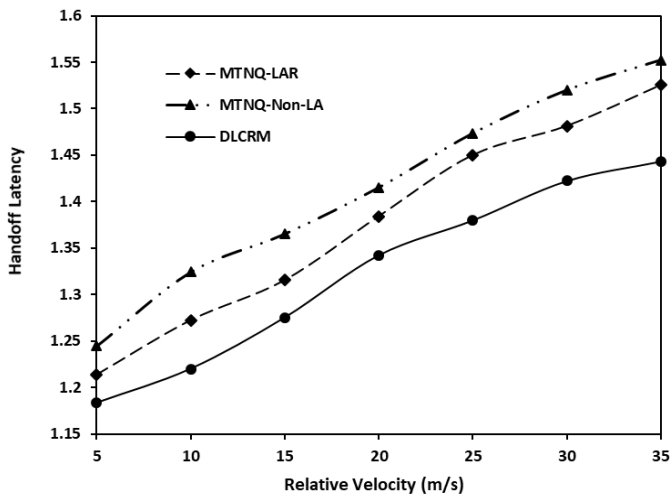

**Fig. 6. Relative Velocity vs Handoff Latency**

# 6  Conclusion

In this paper, deep learning-based channel reservation and channel allocation mechanism for next generation networks is proposed. Various types of calls considered in this paper are real time originating calls, non-real time originating calls, real time handoff calls and non-real time handoff calls. Deep learning concept is used to determine the number of channels that can be exclusively used for specific type of call.

## Conflicts of Interest

Nill.

## Correspondence

D. Sreenivasulu

Research Scholar

Department of Computer Science

Rayalaseema University

Kurnool, AP, India

Email: dsnvas@gmail.com

P. Venkata Krishna

*Senior Member IEEE*

Department of Computer Science

Sri Padmavathi Mahila Visvavidyalayam

Tirupati, India

dr.krishna@ieee.org